\def\papertitle{TIV.lib: an open-source library for the tonal description of musical audio}
\def\paperauthorA{Ant\'onio Ramires}
\def\paperauthorB{Gilberto Bernardes}
\def\paperauthorC{Matthew E. P. Davies}
\def\paperauthorD{Xavier Serra}

\documentclass[twoside,a4paper]{article}
\usepackage{etoolbox}
\usepackage{dafx_20}
\usepackage[draft]{minted}
\usepackage{amsmath,amssymb,amsfonts,amsthm}
\usepackage{euscript}
\usepackage[T1]{fontenc}
\usepackage[utf8]{inputenc}
\usepackage{nimbusserif}
\usepackage{ifpdf}
\usepackage[english]{babel}
\usepackage{caption}
\usepackage{subfig} 
\usepackage{color}
\input glyphtounicode
\ninept

\newcounter{numauth}
\newcounter{listcnt}
\newcommand\authcnt[1]{\ifdefined#1 \stepcounter{numauth} \fi}

\newcommand\addauth[1]{
\ifdefined#1 
\stepcounter{listcnt}
\ifnum \value{listcnt}<\value{numauth}
\appto\authorslist{, #1}
\else
\appto\authorslist{~and~#1}
\fi
\fi}
\authcnt{\paperauthorB}
\authcnt{\paperauthorC}
\authcnt{\paperauthorD}
\authcnt{\paperauthorE}
\authcnt{\paperauthorF}
\authcnt{\paperauthorG}
\authcnt{\paperauthorH}
\authcnt{\paperauthorI}
\authcnt{\paperauthorJ}
\def\authorslist{\paperauthorA}
\addauth{\paperauthorB}
\addauth{\paperauthorC}
\addauth{\paperauthorD}
\addauth{\paperauthorE}
\addauth{\paperauthorF}
\addauth{\paperauthorG}
\addauth{\paperauthorH}
\addauth{\paperauthorI}
\addauth{\paperauthorJ}

\usepackage{times}

\newif\ifpdf
\ifx\pdfoutput\relax
\else
   \ifcase\pdfoutput
      \pdffalse
   \else
      \pdftrue
\fi

\ifpdf 
  \usepackage[pdftex,
    pdftitle={\papertitle},
    pdfauthor={\authorslist},
    pdfsubject={Proceedings of the 23rd International Conference on Digital Audio Effects (DAFx-20)},
    colorlinks=false, 
    bookmarksnumbered, 
    pdfstartview=XYZ 
  ]{hyperref}
  \usepackage[pdftex]{graphicx}
\else 
  \usepackage[dvips]{epsfig,graphicx}
  \usepackage[dvips,
    pdftitle={\papertitle},
    pdfauthor={\authorslist},
    pdfsubject={Proceedings of the 23rd International Conference on Digital Audio Effects (DAFx-20)},
    colorlinks=false, 
    bookmarksnumbered, 
    pdfstartview=XYZ 
  ]{hyperref}
\fi
\usepackage[hypcap=true]{caption}
\title{\papertitle}

\fouraffiliations{
\paperauthorA \,}
{\href{https://www.mtg.upf.edu}{Music Technology Group} \\ Universitat Pompeu Fabra\\ Barcelona, Spain\\
{\tt \href{mailto:antonio.ramires@upf.edu}{antonio.ramires@upf.edu}}
}
{\paperauthorB \,}
{\href{https://www.inesctec.pt/}{INESC TEC and University of Porto}\\ Faculty of Engineering \\ Porto, Portugal \\ {\tt \href{mailto:gba@fe.up.pt}{gba@fe.up.pt}}
}
{\paperauthorC \,}
{\href{https://www.cisuc.uc.pt}{University of Coimbra} \\ CISUC, DEI \\Coimbra, Portugal \\ {\tt \href{mailto:mepdavies@dei.uc.pt}{mepdavies@dei.uc.pt}}
}
{\paperauthorD \,\thanks{\vspace{-3mm}}}
{\href{mtg.upf.edu}{Music Technology Group} \\ Universitat Pompeu Fabra\\ Barcelona, Spain\\ {\tt \href{mailto:xavier.serra@upf.edu}{xavier.serra@upf.edu}}
}

\begin{document}
\ifpdf 
  \DeclareGraphicsExtensions{.png,.jpg,.pdf}
\else  
  \DeclareGraphicsExtensions{.eps}
\fi


\maketitle

\sloppy

\begin{abstract}
In this paper, we present \texttt{TIV.lib}, an open-source library for the content-based tonal description of musical audio signals. Its main novelty relies on the perceptually-inspired Tonal Interval Vector space based on the Discrete Fourier transform, from which multiple instantaneous and global representations, descriptors and metrics are computed---e.g., harmonic change, dissonance, diatonicity, and musical key. The library is cross-platform, implemented in Python and the graphical programming language Pure Data, and can be used in both online and offline scenarios. Of note is its potential for enhanced Music Information Retrieval, where tonal descriptors sit at the core of numerous methods and applications.
\end{abstract}

\section{Introduction}
\label{sec:intro}

In Music Information Retrieval (MIR), several libraries for musical content-based audio analysis, such as Essentia~\cite{bogdanov2013}, Librosa~\cite{mcfee2015librosa}, and madmom~\cite{madmom} have been developed. These libraries have been widely adopted across academia and industry as they promote the fast prototyping of experimental methods and applications ranging from large-scale applications such as audio fingerprinting and music recommendation, to task-specific MIR analysis including chord recognition, structural segmentation, and beat tracking.  

The tonal domain of content-based audio descriptors denotes all attributes related to the vertical (i.e., harmonic) and horizontal (i.e., melodic and voice-leading) combination of tones, as well as their higher-level governing principles, such as the concept of musical key. The earliest research in this domain was driven by the methods applied to symbolic representations of music, e.g., MIDI files. The jump from symbolic to musical audio domain raises significant problems and requires dedicated methods, as polyphonic audio-to-symbolic transcription remains a challenging task~\cite{benetos2018}. While the state of the art~\cite{hawthorne18ismir,ycart2018} in polyphonic music transcription has advanced greatly due to the use of deep neural networks, it remains largely restricted to piano-only recordings. 

One of the most prominent tonal audio descriptors is the chroma vector. This representation divides the energy of the spectrum of an audio signal in the 12 tones of the western chromatic scale across all octaves. This leads to a 12-element vector where each element corresponds to the energy of each pitch class. Throughout this work, this vector will be referred to as the pitch profile. Many algorithms for this representation have been proposed, including Pitch Class Profiles~\cite{Fujishima1999}, Harmonic Pitch Class Profiles (HPCP)~\cite{gomez2006}, the CRP chroma~\cite{muller2010}, and the NNLS chroma ~\cite{matthias2010}. Stemming from this 12-element vector, many metrics and systems have been proposed, for key detection, chord recognition, cover song identification, mood recognition, and harmonic mixing. Yet, despite their fundamental role in many MIR tasks, tonal descriptors are not only less prominent in existing content-based audio libraries, in comparison with rhythmic or timbral descriptors~\cite{bogdanov2013}, but also their perceptual basis is of limited scope~\cite{bernardes2017CMMR-audio}. 

In the context of the aforementioned limitations, we present the~\texttt{TIV.lib}, a cross-platform library for Python and Pure Data, which automatically extracts multiple perceptually-aware tonal descriptions from polyphonic audio signals, without requiring any audio-to-symbolic transcription stage. It owes its conceptual basis to ongoing work within music theory on the DFT of pitch profiles, which has been extended to the audio domain~\cite{bernardes2016JNMR}. The hierarchical nature of the Tonal Interval Vector (TIV) space allows the computation of instantaneous and global tonal descriptors including harmonic change, (intervallic) dissonance, diatonicity, chromaticity, and key, as well as the use of distance metrics to extrapolate different harmonic qualities across tonal hierarchies. Furthermore, it can enable the efficient retrieval of isolated qualities or those resulting from audio mixes 
in large annotated datasets as a simple nearest neighbour-search problem. 

The remainder of this paper is organized as follows. Section~\ref{sec:relatedWork} provides an overview of the ongoing work on pitch profiles DFT-based methods within music theory, followed by a description of the recently proposed TIV space, which extends the method to the audio domain. Section~\ref{sec:architecture} provides a global perspective of the newly proposed \texttt{TIV.lib} architecture. Section~\ref{sec:tiv.lib} details the mathematical and musical interpretation of the description featured in \texttt{TIV.lib} and, finally, Section~\ref{sec:applications} discusses the scope of application scenarios of the library and Section~\ref{sec:conclusions} provides perspectives on future work. 

\section{Related work}
\label{sec:relatedWork}

\subsection{Tonal pitch spaces}
\label{sec:tonalSpaces}
Within the research literature, numerous tonal pitch spaces and pitch distance metrics have been proposed ~\cite{shepard1962,lerdahl2004,tymoczko2010,chew2007}. They aim to capture perceptual musical phenomena by geometrical and algebraic representations, which quantify and (visually) represent pitch proximity. These spaces process pitch as symbolic manifestations, thus capturing musical phenomena under very controlled conditions, with some of the most prominent spaces discarding the pitch height dimension by collapsing all octaves into a 12-tone pitch space.

Attempts to represent musical audio in the aforementioned spaces have been pursued~\cite{chuan2005,haas2008} by adopting an audio-to-symbolic transcription stage. Yet, polyphonic transcription from musical audio remains a challenging task which is prone to error.

Recently, Bernardes et al.~\cite{bernardes2016JNMR} proposed a tonal pitch space which maps chroma vectors derived from audio signals driven into a perceptually inspired DFT space. It expands the aforementioned pitch spaces with strategies to process the timbral/spectral information from musical audio.

\subsection{From the DFT of symbolic pitch distributions to the Tonal Interval Vector space}

In music theory, the work proposed by Quinn~\cite{quinn2007} and followed by~\cite{amiot2016,yust2019,tymoczko2019} on the Discrete Fourier Transform (DFT) of pitch profiles, has been shown to elicit many properties with music-theoretic value. Moreover, in~\cite{Dawson2020} DFT-based pitch spaces were shown to capture human perceptual principles. 

In the Fourier space, a $6$-element complex vector, corresponding to the $1 \leq k \leq 6$ DFT coefficients, is typically adopted. The magnitude of the Fourier coefficients has been used to study the shape of pitch profiles, notably concerning the distribution of their interval content. This allows, for example, to quantify diatonic or chromatic structure (see Section~\ref{sec:tiv.lib} for a comprehensive review of the interpretations of the coefficients). The phase of the pitch profiles in the Fourier space reveals aspects of tonal music in terms of voice-leading~\cite{tymoczko2008}, tonal regions modelling and relations~\cite{yust2017}, and the study of tuning systems~\cite{amiot2016}. In summary, the magnitude of the pitch profiles express harmonic quality and the phases harmonic proximity.

Recently, a perceptually-inspired equal-tempered, enharmonic, DFT-based TIV space~\cite{bernardes2016JNMR} was proposed. One novelty introduced by this newly proposed space in relation to remaining Fourier spaces was the combined use of the six coefficients in a TIV, $T(k)$. Moreover, the perceptual basis of the space is guaranteed by weighting each coefficients by empirical ratings of dyad consonance, $w_a(k)$. $T(k)$ allows the representation of hierarchical or multi-level pitch due to the imposed $L_1$ norm, such that:
\begin{equation}
\begin{split}
T(k)= w_a(k) \sum_{n=0}^{N-1} \bar{c}(n) e^\frac{-j2\pi  kn}{N} \, ,\\
k \in  \mathbb{Z} \quad \textrm{with} \quad  \bar{c}(n)=\frac{c(n)}{\sum_{n=0}^{N-1}c(n)}
\label{eq.1}
\end{split}
\end{equation}
where $N$=$12$ is the dimension of the chroma vector, $c(n)$, and $k$ is set to $1 \leq k \leq 6$ for $T(k)$ since the remaining coefficients are symmetric. The weights, $w_a(k)=\{3, 8, 11.5, 15, 14.5, 7.5\}$, adjust the contribution of each dimension $k$ of the space to comply with empirical ratings of dyad consonance as summarised in~\cite{huron1994interval}. $w_a(k)$ accounts for the harmonic structure of musical audio driven from an average spectrum of orchestral instruments~\cite{bernardes2017CMMR-audio}.

\section{TIV.LIB: Implementation}
\label{sec:architecture}

The \texttt{TIV.lib} includes several signal-processing functions or descriptors for characterising the tonal content of musical audio. This library is implemented in Python, using only \texttt{Numpy} and \texttt{Scipy} as dependencies and Pure Data, with both available to download at: \url{http://bit.ly/2pBYhqZ}. Illustrative analysis of musical audio examples for the descriptors are provided in the Python download link as a Jupyter Notebook. The Python implementation targets batch offline processing and the Pure Data implementation online processing. 

As an input, the library takes $12$-element chroma vectors, $c(n)$, from which TIVs, $T(k)$, are then computed.~\footnote{A tutorial example on the extraction of HPCP representations from audio is provided in the library package, both using Essentia and Librosa.} Any input representation will have an effect on the space, as such we leave the choice of which chroma representation up to the user in order to best fit the problem at hand. Although the system is agnostic to the chosen chroma, we recommend the 
``cleanest'' chroma representation, i.e., that which is closest to a symbolic representation, to be selected. The time scale of the TIV is dependent of the adopted window size during the chroma vector computation. For instantaneous TIVs, a single-window chroma vector can be used as input. For global TIVs, consecutive chroma vectors can be averaged across the time axis prior to the TIV computation.

\begin{figure}
  \begin{center}
  \includegraphics[width=230pt]{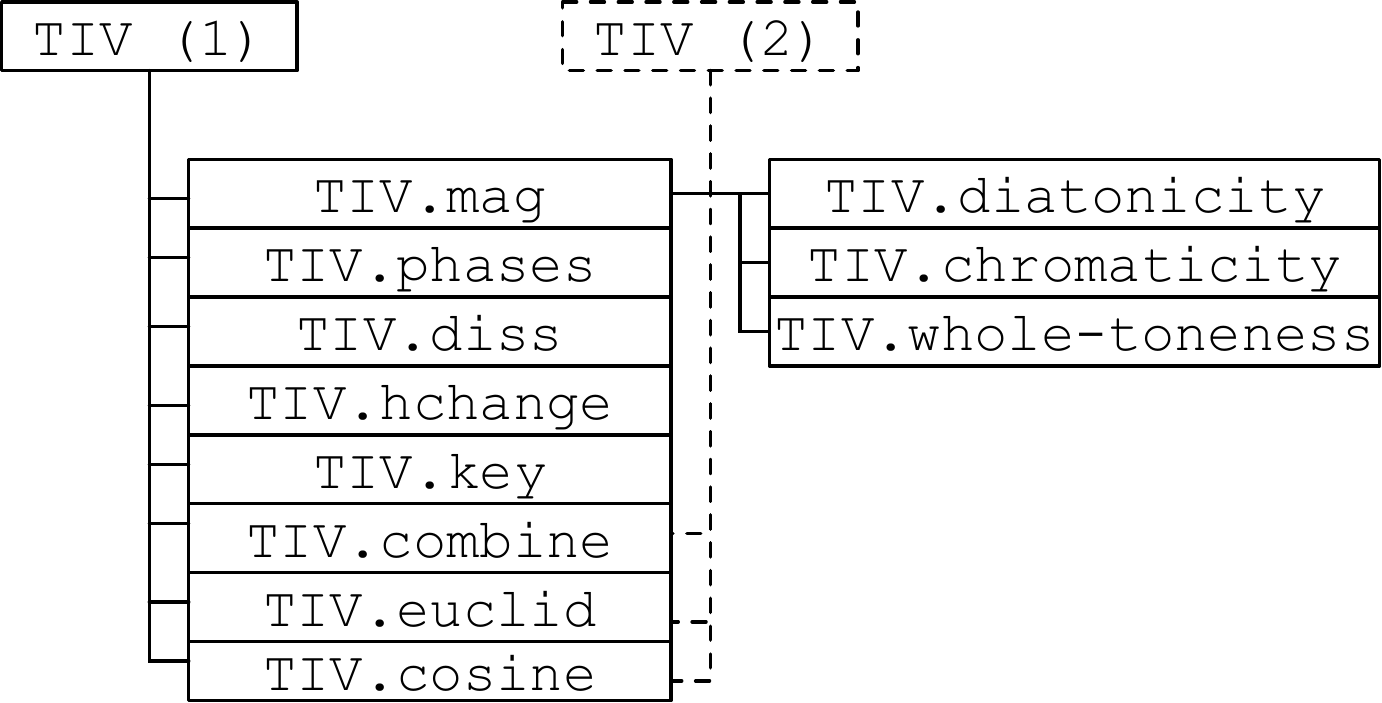}
  \end{center}
  \caption{A graph of the dependencies of the feature extraction modules of \texttt{TIV.lib}. The algorithms connected to TIV(2) through a dashed line require two inputs for the feature calculation.}
  \label{fig:arch}
\end{figure}

In Figure \ref{fig:arch} we present the architecture of \texttt{TIV.lib}. In this graph of dependencies we can see the algorithms that have been implemented and which classes they require for their calculation. 

\section{TIV.lib: Algorithms}
\label{sec:tiv.lib}

This section details the functions included in the \texttt{TIV.lib}, focusing on their mathematical definition and musical interpretation.

\textbf{\texttt{TIV}} is a 
$6$-element complex vector, which transforms chroma into an interval vector space by applying Eq.~\ref{eq.1}, an $L_1$-norm weighted DFT. The resulting space dimensions combine intervallic information in the coefficients' magnitude and the tonal region (i.e., musical key area) it occupies in the coefficients' phase. The mapping between chroma and the TIV retains the bijective property of the DFT and allows the representation of any variable-density pitch profile in the chroma space as a unique location in the TIV space.

\textbf{\texttt{TIV.mag}} is a $6$-element (real) vector that reports the magnitude of the TIV elements $1 \leq k \leq 6$, such that:
\begin{equation}
mag(k) = ||T(k)||
\end{equation}
It provides a characterisation of the harmonic quality of a pitch profile, namely its intervallic content, distilling the same information as the pitch-class interval vector \cite{forte1964,forte1973}. Mathematically, it is well-understood that a large magnitude in $T(k)$ coefficients indicates how evenly the pitch profile can be divided by $N/k$. Musically, the work on the DFT of pitch profiles \cite{bernardes2016JNMR,quinn2007} emphasizes the association between the magnitude of Fourier coefficients and tonal qualities:
$||T(1)|| \leftrightarrow chromaticity$, $||T(2)|| \leftrightarrow dyadicity$, $||T(3)|| \leftrightarrow triadicity$,
$||T(4)|| \leftrightarrow diminished \, quality$,
$||T(5)|| \leftrightarrow diatonicity$,
$||T(6)|| \leftrightarrow whole-toneness$. Please refer to \cite{amiot2016, yust2019} for a comprehensive discussion on the interpretation of the DFT coefficients.\footnote{We note for each of these single Fourier coefficient quantities that the effects of the weights can be factored out.} One distinct property of the \texttt{TIV.mag} vector is its invariance under transposition or inversion~\cite{amiot2016}. For example, all major triads or harmonic minor scales share the same Fourier magnitude, hence the same \texttt{TIV.mag} vector~\footnote{Note that the phases, discarded here, will differ. As such, the uniqueness property of the TIV is maintained as it combines both magnitude and phase information.}.

\textbf{\texttt{TIV.phases}} is a $6$-element (real) vector that reports the phases (or direction) of the TIV coefficients $1<k<6$, such that:
\begin{equation}
phases(k) = \angle T(k)
\end{equation}
It indicates which of the transpositions of a pitch profile quality is under analysis \cite{hoffman2008}, as transposition of a pitch profile by $p$ semitones, i.e., circular rotations of the chroma, $c(n)$, rotates the $T(k)$ by $\varphi(p)=\frac{-2{\pi}kp}{N}$. TIV phases are also associated with regional (or key) areas, whose diatonic set is organised as clusters in the TIV space \cite{bernardes2016JNMR,yust2017}.

\textbf{\texttt{TIV.combine}} computes the resulting TIV from mixing (or summing) multiple TIVs representing different musical audio signals. Due to the properties of the DFT space, this operation can be efficiently computed as a linear combination of any number of TIVs, $T(k)$. Given TIVs $T_1(k)$ and $T_2(k)$, their linear combination, weighted by their respective energy, $a_1$ and $a_2$, is given by:
\begin{equation}
T_{1+2}(k)=\frac{T_1(k) \cdot a_1 + T_2(k) \cdot a_2} {a_1+a_2} 
\end{equation}
$a_1$ and $a_2$ are retrieved from the discarded DC components $T_1(0)$ and $T_2(0)$. 

\begin{figure}
  \includegraphics[width=\columnwidth]{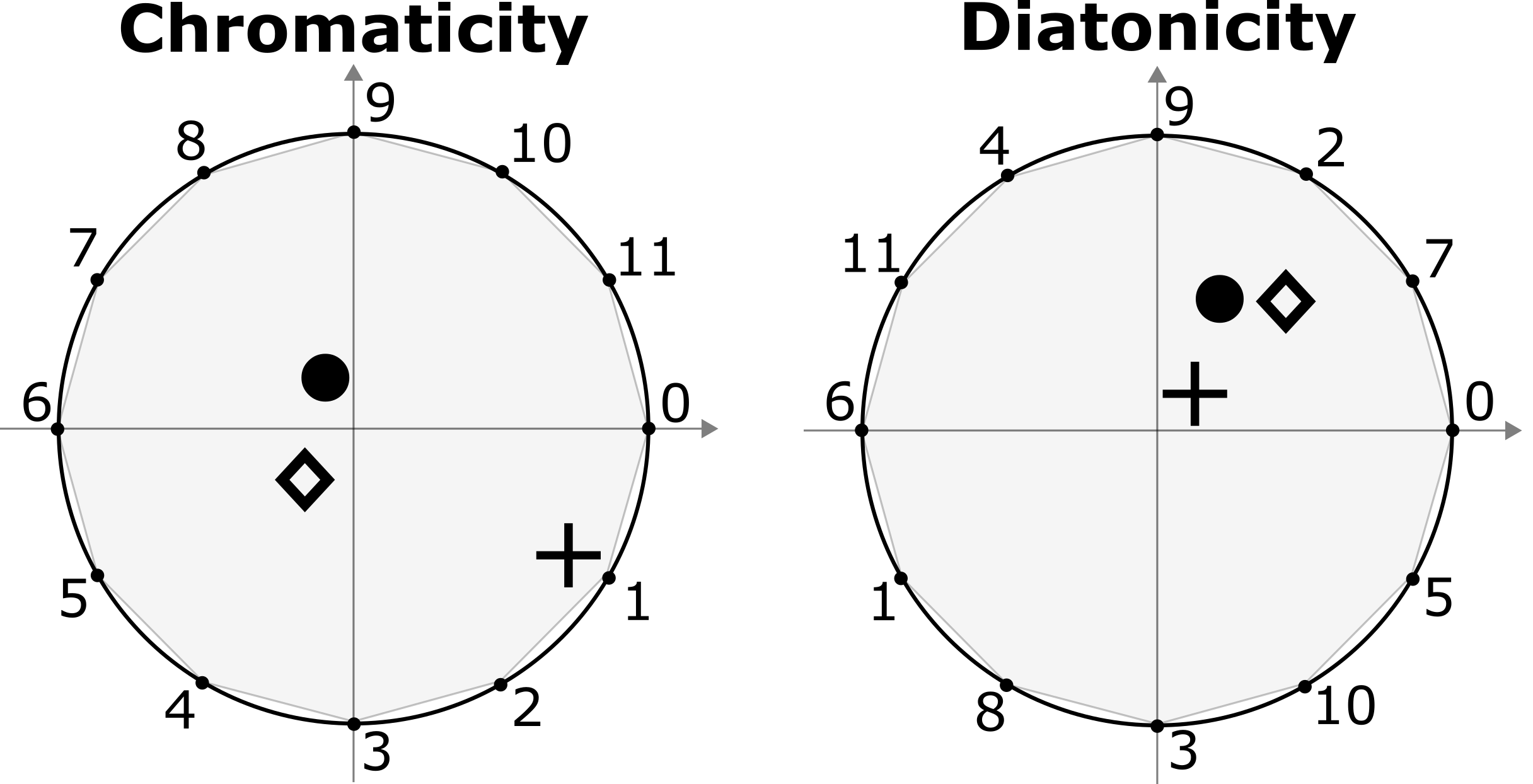}
  \caption{Two DFT coefficients interpreted as chromaticity and diatonicy. Three TIV are plotted for comparison: C major chord \{0,4,7\} (\boldmath{$\diamondsuit$}), 3-note chromatic cluster \{0,1,2\} (\textbf{+}), and C major scale \{0,2,4,5,7,9,11\} (\boldmath{$\bullet$}). The grey shaded areas indicate the space the TIVs can occupy.}
  \label{fig:TIV}
\end{figure}

\textbf{\texttt{TIV.chromaticity}} reports the level of concentration of a sonority in a specific location of the chromatic pitch circle as a value within the [$0$,$1$] range, computed as the magnitude of the $T(1)$ normalized to unity: $\frac{||T(1)||}{w_a(1)}$. This value is close to $0$ for sounds exhibiting energy in evenly-spaced pitch classes (such as typically tonal chords and scales) and close to $1$ for chromatic pitch aggregates. 

\textbf{\texttt{TIV.diatonicity}} reports the level of concentration of a sonority within the circle of fifths as a value within the [$0$,$1$] range. The larger the magnitude of the $T(5)$ normalized to unity, $\frac{||T(5)||}{w_a(5)}$, the higher the level of diatonicity.

\textbf{\texttt{TIV.whole-toneness}} reports the proximity to one of the two existing whole-tone collection within the $12$-tone equal temperament tuning. The level of whole-toneness is reported within the [$0$,$1$] range resulting from the magnitude of the $T(6)$ normalized to unity, such that: $\frac{||T(6)||}{w_a(6)}$. 

Fig.~\ref{fig:TIV} shows the DFT coefficients from which we extract chromaticity and diatonicity descriptions as the magnitude of $T(1)$ and $T(5)$, respectively. We plot pitch profiles that aim to illustrate the behaviour of each coefficient in eliciting the chromatic and diatonic character of the C major chord and C major scale as well as chromatic 3-tone cluster by inspecting their magnitude. Note that the magnitude of both the C major chord and C major scale, two prototypical diatonic pitch profiles, clearly have greater magnitude in the diatonic $T(5)$ coefficient in comparison with the three-note cluster, a prototypical chromatic profile. Conversely, in the chromatic $T(1)$ coefficient, the magnitudes of the above pitch profiles show the expected opposite behaviour, thus mapping the three-note cluster further from the centre.            

\textbf{\texttt{TIV.euclid}} and \textbf{\texttt{TIV.cosine}} compute the Euclidean, $E$, and cosine, $C$, distance between two given TIVs, $T_1(k)$ and $T_2(k)$, using Eqs.~\ref{eq.euclid} and \ref{eq.cosine}, respectively.

\begin{equation}
E\{T_1, T_2\}=\sqrt{||T_1-T_2||^2}
\label{eq.euclid}
\end{equation}

\begin{equation}
C\{T_1, T_2\}=\frac{T_1 \cdot T_2}{||T_1||||T_2||}
\label{eq.cosine}
\end{equation}

The cosine distance (i.e., the angular distance) between TIVs can be used as an indicator of how well pitch profiles ``fit" or mix together. For example, it quantifies the degree of tonal proximity of TIV mixtures, or informs which translation or transposition of a TIV best aligns with a given key. Conversely, Euclidean distances between TIVs relate mostly to melodic (or horizontal) distance. It captures the neighbouring relations observed in the \textit{Tonnetz}, where smaller distances agree with parsimonious movements between pitch profiles. Please refer to~\cite{tymoczko2019,tymoczko2008} for a comprehensive discussion on this topic.

\textbf{\texttt{TIV.hchange}} computes a harmonic change detection function across the temporal dimension of an audio signal. Peaks in this function indicate transitions between regions that are harmonically stable. We compute a harmonic change measure, $ \lambda $, for an audio frame $m$ as the Euclidean distance between frames $m+1$ and $m-1$ (Eq.~\ref{tis.hchange}), an approach inspired by Harte et al.~\cite{harte2006}, which can be understood as adopting three coefficients out of the $1 \leq k \leq 6$ of the TIV, $T(k)$, i.e., those corresponding to the circle of fifths, the circle of minor thirds, and the circle of major thirds.

\begin{equation}
\lambda_m=\sqrt{ || T_{m-1}-T_{m+1} ||^2}
\label{tis.hchange}
\end{equation}

\textbf{\texttt{TIV.diss}} provides an indicator of (interval content) dissonance, as the normalized TIV magnitude subtracted from unity, $1-\frac{|T(k)|}{|w_a(k)|}$. This perceptually-inspired indicator stems from the weighted magnitude of the TIV coefficients, which rank the intervals $1\leq k \leq 6$ to match empirical ratings of dissonance within the Western tonal music context~\cite{bernardes2017CMMR-audio,bernardes2016JNMR}.

\textbf{\texttt{TIV.key}} infers the key from an audio signal as a pitch class (tonic) and a mode (major or minor). It is computed as the Euclidean distance from the $24$ major and minor key TIVs, $T^{p\star}_r(k)$, defined as the shifts (i.e. rotation) of the $12$ major and $12$ minor profiles, $p$, by Temperley~\cite{temperley1999} or Sha'ath~\cite{sha2011}, such that: 
\begin{equation}
    R_{min} = \textrm{argmin}_r \sqrt{ \vert \vert T \cdot \alpha - T^{p\star}_r\vert \vert ^2}
\end{equation}
where $T^{p\star}_r$ are $24$ major and minor key profiles TIVs, $p$. When $r \leq 11$ , we adopt the major profile and when $r \geq 12$, the minor profile. $\alpha$ is a bias introduced to balance the distance between major and minor keys. Optimal values of $\alpha=0.2$ and $\alpha=0.55$ have been proposed in~\cite{bernardes2017ICASSP} for the Temperley~\cite{temperley1999} and Shat'ath~\cite{sha2011} key profiles, respectively. The output is an integer, $R_{min}$, ranging between $0-11$ for major keys and $12-23$ for minor keys, where $0$ corresponds to C major, $1$ to C\# major, and so on through to $23$ being B minor.

\section{Applications and Perspectives}
\label{sec:applications}
Following the emerging body of music theory literature on the DFT of pitch profiles and the continuous work of the TIV space, we implemented the perceptually-inspired \texttt{TIV.lib} in Python and Pure Data. The former aims at batch offline processing and the latter mostly at online or real-time processing, but allowing offline computations as well.

\begin{figure}
  \begin{center}
  \includegraphics[height=60mm]{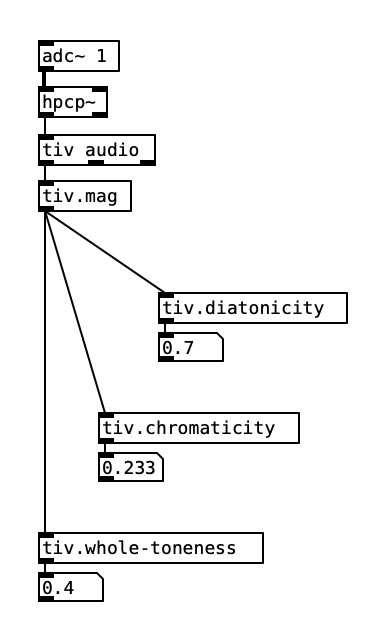}
  \end{center}
  \caption{Pure Data patch for online computation of diatonicity, chromaticity and whole-toneness harmonic qualities of a live input audio signal, using the \texttt{TIV.lib}.}
  \label{fig:pd}
\end{figure}

Figure~\ref{fig:pd} shows an example usage of the \texttt{TIV.lib} functions for computing the diatonicity, chromaticity and whole-toneness harmonic qualities in Pure Data. 

\begin{figure}[ht]
  \begin{center}
  \includegraphics[height=55mm]{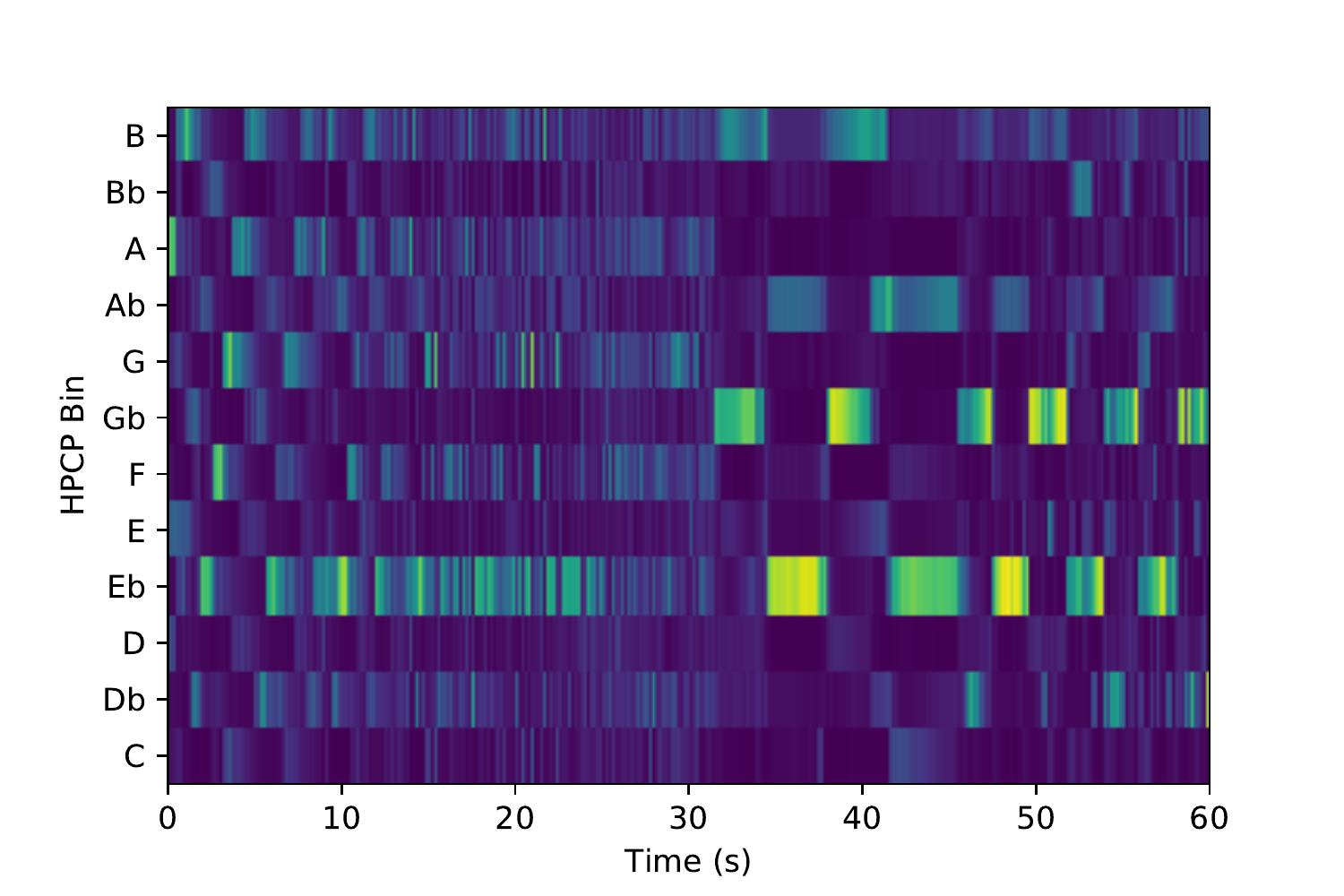}
  \end{center}
  \caption{Chromagram of the first minute of Kraftwerk's ``Spacelab.''}
  \label{fig:kraftChroma}
\end{figure}

\begin{figure}[ht]
  \begin{center}
  \includegraphics[height=55mm]{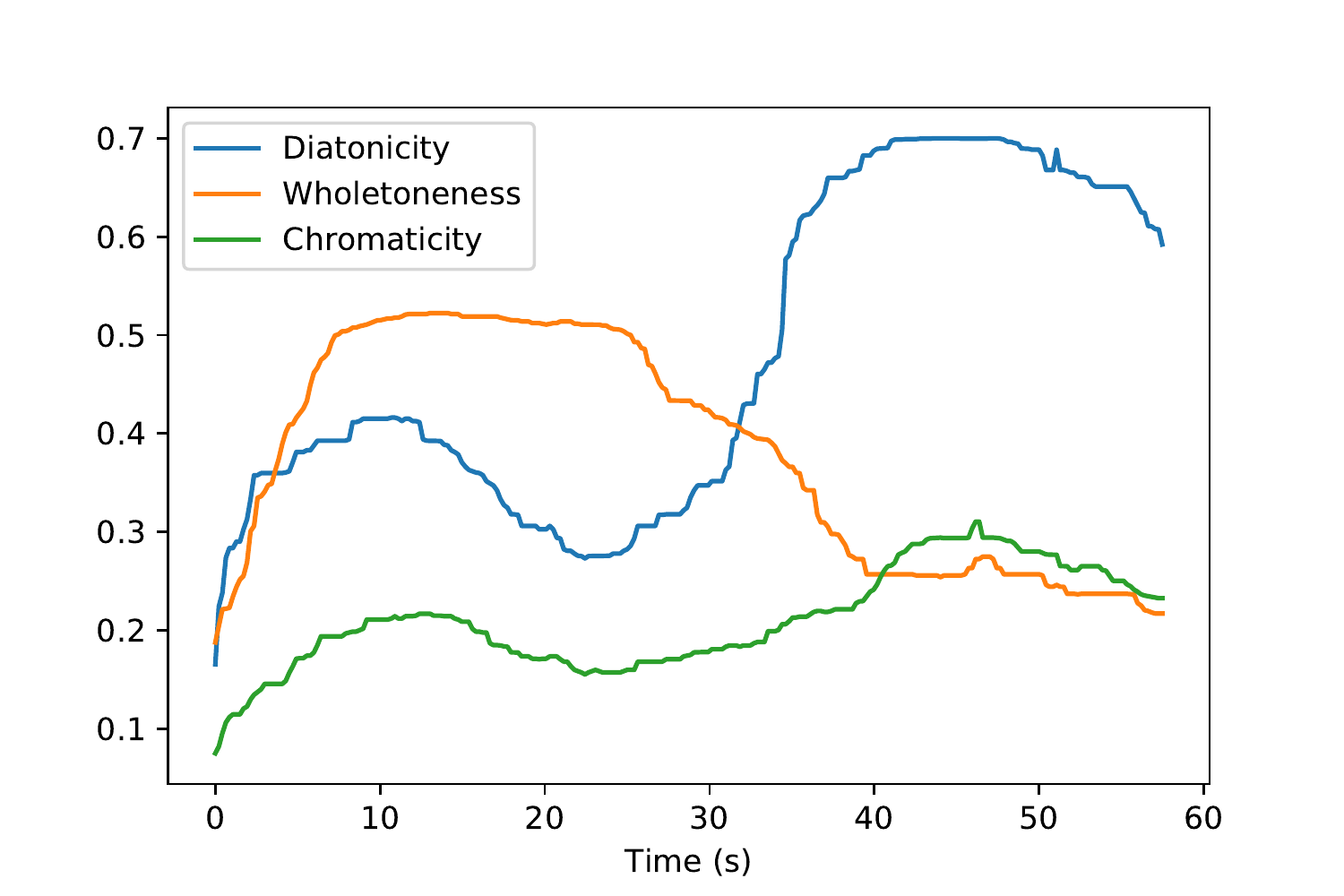}
  \end{center}
  \caption{Output of the diatonicity, chromaticity and whole-toneness harmonic qualities the first minute of Kraftwerk's ``Spacelab'', using the \texttt{TIV.lib}.}
  \label{fig:kraft}
\end{figure}

In order to achieve the same result in Python, the following code can be executed:

\begin{minted}{python}
import TIVlib as tiv

ex_tiv = tiv.TIV.from_pcp(example_chroma)
ex_wholetoneness = ex_tiv.wholetoneness()
ex_diatonicity = ex_tiv.diatonicity()
ex_chromacity = ex_tiv.chromaticity()
\end{minted}

We now provide an example usage of this library for extracting tonal features of a musical piece. We run this code for an excerpt of the Kraftwerk song ``Spacelab,'' to demonstrate how this library can provide useful information related to its diatonicity and whole-toneness. As can be seen from the Chromagram in Figure~\ref{fig:kraftChroma}, this song starts in the whole tone scale [F\# G\# A\#C D E], and then moves, at $33$\,s, to a diatonic set [C D Eb F G Ab Bb C]. In Figure~\ref{fig:kraft} we show this by plotting the evolution of the\texttt{TIV.diatonicity}, \texttt{TIV.wholetoneness} and \texttt{TIV.chromaticity} outputs for this music. 

Several functions of the \texttt{TIV.lib} result from ongoing research and have been evaluated in previous literature~\cite{bernardes2017CMMR-audio, bernardes2016JNMR, bernardes2017ICASSP, bernardes2016ACM}, where the possibility of the TIV space to geometrically and algebraically capture existing spaces of perceptual and music theoretical value, such as Euler and Krumhansl~\cite{bernardes2016JNMR}, were shown. In particular, we highlight our previous work on key recognition~\cite{bernardes2017ICASSP} and harmonic mixing~\cite{bernardes2017CMMR-audio,macas2019}, where TIV-based approaches outperformed more traditionally used harmonic features. In addition, the use of TIVs can also extend content-based audio processing by providing a vector space where distances and metrics (e.g., dissonance and harmonic proximity) among multi-level pitch, chords, and keys, capture perceptual aspects of musical phenomena. Examples of creative possibilities of the TIV space have also been shown in Musikverb~\cite{musikverb}, where it was used for developing a novel type of harmonically-adaptive reverb effect.

We strongly believe that the properties of the TIV space can be further explored in content-based audio processing. For example, the possibility to isolate the harmonic quality in \texttt{TIV.mag} as a pitch-invariant audio representation can be relevant for several MIR tasks that rely on multiple transposed versions of a given musical pattern, such as in query-by-humming, and cover song detection. Moreover, the possibility to compute TIV mixes as a computationally efficient linear combination allows for the fast retrieval of musical audio from large datasets (e.g., Freesound~\cite{Font2013}), as a simple nearest-neighbour search problem. Finally, the newly proposed indicators of tonal quality such as \texttt{TIV.chromaticity}, \texttt{TIV.diatonicity}, \texttt{TIV.wholte-toneness}, and \texttt{TIV.diss} not only extend musical theoretical methodologies to content-based processing from audio performance data, but can also promote a greater understanding of tonal content in MIR tasks.

By providing streamlined access to a set of music theoretic properties which are non-trivial to obtain from commonly used time-frequency representations in MIR such as the STFT (or even from chroma-like representations directly), we believe the \texttt{TIV.lib} can lay the foundation for a kind of ``enhanced'' MIR in tasks such as chord recognition and key estimation which can directly leverage the complementary contextual information contained within the \texttt{TIV.lib} descriptors.

\section{Conclusions}
\label{sec:conclusions}
In this paper we have introduced the open-source tool, \texttt{TIV.lib}, as a means to drive the uptake and usage of the Tonal Interval Space both in offline music signal analysis via the python implementation, as well as in online contexts using Pure Data. While we hope to see a growth of applications which benefit from access to music theoretic harmonic features provided by \texttt{TIV.lib} our own future work will focus in two principal areas: i) investigating the processing stages which directly precede the calculation of the TIV; and ii) in the application of the TIV across large datasets. More specifically, we seek to study the impact of different methods for calculating the requisite chroma vectors (e.g., HPCP~\cite{gomez2006},  NNLS~\cite{matthias2010}, or timbre-invariant chroma~\cite{muller2010}) in the TIV space, as pursued in ~\cite{bernardes2017ICASSP, bernardes2016ACM} within the scope of audio key detection. Furthermore, we will study an optimal strategy to define the weights, $w_a(k)$, for particular audio sources and to implement the descriptors in a large online musical database supported by content-based analysis, as a strategy to study the descriptors under a large-scale environment for musical retrieval and creation. Finally, we intend to add this library to existing musical audio analysis libraries such as Essentia and Librosa.

\section{Acknowledgments}
Ant\'onio Ramires is supported by the European Union’s Horizon 2020 research and innovation programme under the Marie Skłodowska-Curie grant agreement No765068, MIP-Frontiers.

Gilberto Bernardes is supported by Experimentation in music in Portuguese culture: History, contexts, and practices in the 20th and 21st centuries—Project co-funded by the European Union, through the Operational Program Competitiveness and Internationalization, in its ERDF component, and by national funds, through the Portuguese Foundation for Science and Technology.

This work is funded by national funds through the FCT - Foundation for Science and Technology, I.P., within the scope of the project CISUC - UID/CEC/00326/2020 and by European Social Fund, through the Regional Operational Program Centro 2020, as well as by Portuguese National Funds through the FCT - Foundation for Science and Technology, I.P., under the project IF/01566/2015.

\nocite{*}
\bibliographystyle{IEEEbib}
\bibliography{DAFx20_tmpl} 

\end{document}